\documentstyle[prb,aps,eqsecnum,multicol,psfig,array]{revtex}
\draft
\tighten
\begin{document}
\title{Charging spectrum of a small Wigner crystal island.}
\author{A.~ A.~ Koulakov and B.~ I.~ Shklovskii}
\address{Theoretical Physics Institute, University of Minnesota,
Minneapolis, Minnesota 55455}

\date{\today}

\maketitle
\begin{abstract}

Charging of a clean two-dimensional island is studied 
in the regime of small concentration of electrons when they form the Wigner crystal.
The number of electrons in the island is assumed to be not too big ($N \le 100$).
It is shown that the total energy of the island as a function  of $N$
has a quasi-periodic component of 
a universal shape, that is independent of the form of 
electron-electron interactions.
These oscillations are caused by the combination of the geometric effects
associated with packing of the triangular lattice into the circular
island. These effects are: the shell effect, associated with starting a 
new crystalline row, and the so-called confinement 
polaronic effect. 
In the presence of close metallic gates, which eliminate the long-range
part of the electron-electron interactions, the oscillations of the energy
bring about simultaneous entering of the dot by a few electrons.  
\end{abstract}
\begin{multicols}{2}

% ---------------------------------------------------------------------------
%
%	Introduction
%	
%
%
% ---------------------------------------------------------------------------

\section{Introduction}
\label{Introduction}

In recent experiments\cite{AshooriNew,AshooriOld}
the charging of a quantum dot is studied by the single
electron capacitance spectroscopy method. The quantum dot is located
between two capacitor plates: metallic gate and heavily doped GaAs layer.
Tunneling between the dot and the heavily doped side is possible during
the experimental times while barrier to the metal is completely insulating.
DC potential $V_g$ and a weak AC potential are applied to the
capacitor. With the increase of $V_g$ the differential capacitance experiences
periodic peaks when addition of a new electron to the dot  becomes possible.
The spacing between two nearest peaks $\Delta V_{g}$ can be related to the ground state
energy $E(N)$ of the dot with $N$ electrons:
\begin{equation}
\begin{array}{ll}
{\displaystyle
\alpha e \Delta V_{g}}&{\displaystyle=E(N+1) - 2E(N) + E(N-1)
} \\ \\
{}&{\displaystyle =\Delta(N) \equiv e^{2}/C_{N}.}
\end{array}
\label{capacitance}
\end{equation}
Here $\alpha$ is a geometrical coefficient, 
$\Delta(N)$ is the charging energy, $C_{N}$ is
the capacitance of the dot with $N$ electrons. 
It was observed in Refs.~\onlinecite{AshooriNew,AshooriOld}
that at a low concentration of electrons or in a strong magnetic field
the nearest peaks can merge, indicating that at some values of $V_{g}$ 
two electrons enter the dot simultaneously. 
In other words some charging energies apparently become zero or negative.
In a fixed magnetic field this puzzling event
is repeated periodically in $N$. Disappearance of the charging energy looks
like a result of unknown attraction between electrons and represents a
real challenge for theory.

Explanation of the pairing of the differential capacitance peaks 
based on the {\em lattice}
polaronic mechanism has been previously suggested in  Ref.\onlinecite{Phylipps}.
In Ref.\onlinecite{Raikh} it was demonstrated how
electron-electron repulsion screened by a close metallic gate  
can lead to electron pairing 
for a specially arranged compact clusters of localized states
in a {\em disordered} dot.
This effect is a result of redistribution of the other electrons after 
arrival of new ones. It was interpreted in Ref.\onlinecite{Raikh}
as {\em electronic bipolaron}.

In this paper we study the addition spectrum of a dot where density of 
electrons is small and disorder is weak enough for electrons
to form the Wigner crystal.  We call such a dot
the Wigner crystal island. In the experimental conditions of 
Ref.\onlinecite{AshooriNew} 
one can think about Wigner crystal island literally
only in the highest magnetic field. One can also imagine
similar experiments with a Wigner crystal island on the surface of liquid helium.

We show that when a metallic gate is very close to the plane of 
a two-dimensional (2D) island, 
two or more addition levels can be anomalously close and even merge similar 
to the observations of Ref.\onlinecite{AshooriNew}.
This phenomenon repeats periodically,
and is associated with the formation of highly 
symmetric electronic configurations, 
consisting of an integer number of closed radial shells.
The new electrons added to such a configuration reside on its surface.
They exert some force onto the symmetric configuration making it move 
in the external confinement field.
This displacement facilitates the entrance of the new electrons
mediating an effective attraction between them.
To emphasize that this effect is produced by the confinement potential
and to stress its multi-electron nature we call it
{\em confinement multi-polaronic} effect.

The paper is organized as follows. In Section~\ref{Numerical_Simulations}
we present our model and the results of its numerical solution.
In Sections ~\ref{Toy_Model} and ~\ref{Coulomb} we give the explanations of
the numerical results for the screened and for the unscreened Coulomb interactions
respectively. Section ~\ref{Conclusion} is dedicated to our conclusions.

% ---------------------------------------------------------------------------
%
%	Numerical Simulations
%	
%
%
% ---------------------------------------------------------------------------

\section{Numerical Simulations}
\label{Numerical_Simulations}

In this Section we consider the system of $N$ interacting electrons 
in the parabolic confinement. 
The energy of the system is given by the following expression:
\begin{equation}
E = \sum_{i < j} {U\left( \bbox{r}_i - \bbox{r}_j \right) } + A \sum_i \bbox{r}_i^2.
\label{Energy}
\end{equation}
Here $A$ plays the role of strength of the confinement, and
$U\left( \bbox{r}\right)$ is the interaction potential.
The forms of the interaction potential considered are:
the pure Coulomb interaction
\begin{equation}
U\left( \bbox{r}\right) = {e^2} / {\kappa r},
\label{Coulomb_int}
\end{equation}
with $e$ and $\kappa$ being
the electron charge and the dielectric constant correspondingly,
and the exponential interaction
\begin{equation}
U\left( \bbox{r}\right) = U_0\exp \left( -r/d \right).
\label{Exp_int}
\end{equation}
The latter interaction potential corresponds to the case
when the island is situated between two metallic gates,
with $\pi d$ being the distance between them and
$U_0 \sim e^2/\kappa d$.

This energy was minimized over the coordinates of electrons using
the numerical technique similar to that outlined in Ref.~\onlinecite{Peeters94}.
It is the standard Metropolis simulated annealing algorithm. 
The temperature of the simulation was decreased exponentially (quenched)
in $10^5N$ simulation steps. The average displacement of electrons
in a step was chosen automatically to assure the acceptance  
probability of a new configuration of $0.3$.
Each time the simulation for the new number of electrons was started
from a random set of coordinates. 
We used a few restarts for a given number of electrons.
Although the configurations obtained in different restarts 
sometimes were different, the maximum difference in energy is
typically $10$ times smaller than the systematic variations of $E(N)$. 
Thus although we cannot guarantee that we have found the ground
state of the electron system, we are confident that these 
variation of $E\left( N\right)$ are reproduced reliably.
Our results for energies agree with those of Ref.~\onlinecite{Peeters94}
and for some $N$ are even lower. 
Before presenting these results  
we would like to discuss the method of the data processing.

\begin{figure}
\centerline{
\psfig{file=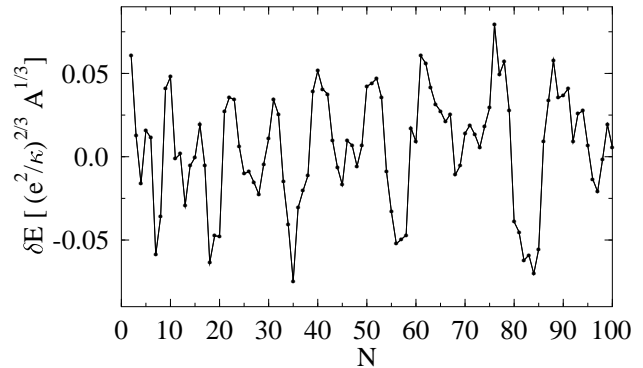,height=1.9in,bbllx=50pt,bblly=104pt,bburx=490pt,bbury=369pt}
}
\vspace{0.1in} 
\setlength{\columnwidth}{3.2in}
\centerline{\caption{
Fluctuating part of energy of the Wigner crystal
island. 
\label{fig1}
}}
\vspace{-0.1in}
\end{figure}

\begin{figure}
\centerline{
\psfig{file=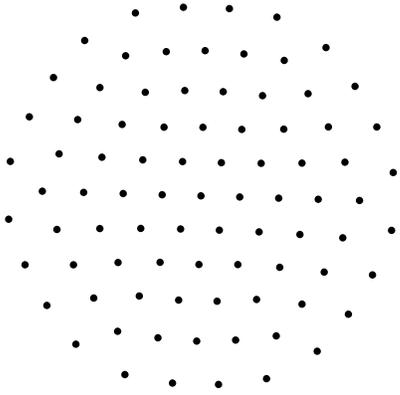,height=2.0in,bbllx=46pt,bblly=107pt,bburx=350pt,bbury=401pt}
}
\vspace{0.1in} 
\setlength{\columnwidth}{3.2in}
\centerline{\caption{
The magic number configuration \protect{$N=85$}.
\label{fig2}
}}
\vspace{-0.1in}
\end{figure}

Consider first the problem in the case of no screening, that is 
described in our model by Eqs. ~(\ref{Energy}) and (\ref{Coulomb_int}).
The instructiveness of this problem is in the
fact that the corresponding electrostatic problem (ignoring the discreteness 
of the charge of electron) can be solved exactly~\cite{Sneddon66}.
The solution for the density of electrons can be shown to be hemisphere:
\begin{equation}
\begin{array}{l}
{\displaystyle
n\left( r\right) = n_0\sqrt{1- \frac {r^2}{R^2}}} \\ \\
{\displaystyle
n_0 = \frac {4A\kappa R}{\pi^2e^2} ,~
R = \left( \frac {3\pi Ne^2}{8A\kappa} \right)^{1/3}.
}
\end{array}
\label{n_r}
\end{equation}
The total potential of the electrons is equal to $\phi_0 = 2AR^2$.
This solution allows us to calculate the scaling of all other contributions
to the energy of the ground state with the number of particles 
(see Appendix~\ref{App_A}). In particular:
\begin{equation}
\begin{array}{ll}
{\displaystyle
\bar{E}\left( N \right) = } & 
{ \left( e^2/\kappa \right) ^{2/3} A^{1/3} \left( \displaystyle \eta_1 N^{5/3}
+ \right.} \\ 
 {~} & {\left. \displaystyle + 
\eta_2 N^{7/6} + \eta_3 N^{2/3} + 
\eta_4 N^{7/15} + \ldots \right)} ,
\end{array}
\label{E_ave}
\end{equation}
where $\eta_i$ are some constants.
The first term in this asymptotic series is the electrostatic
energy that can be found from the solution cited above.
The next three term in the order of appearance are:
the correlation energy, the overscreening energy associated 
with the screening of the external potential by the Wigner crystal, 
and the surface energy. The coefficients $\eta_i$ can be
found from the best fit to the numerical data:
$\eta_1 = 6/5 \left( {3\pi}/8 \right)^{2/3}$ , 
$\eta_2 = -1.0973(2)$ , $\eta_3 = -0.182(3)$ , $\eta_4 = -0.06(1)$.
This smooth contribution to energy can now be subtracted from the 
numerical data to obtain the fluctuating part:
\begin{equation}
\delta E \left( N\right) = E \left( N\right) - \bar{E} \left( N\right).
\label{dE}
\end{equation}
The fluctuating part is displayed in Fig.~\ref{fig1}.
%---------------------------------------------------------------------------------
%
%	Fig.1
%
%---------------------------------------------------------------------------------
The curve in Fig.~\ref{fig1} evidently has a quasi-periodic structure.
It consists of the series of interchanging deep and shallow 
minima, separated by the peeks with more or less 
smooth slopes.
We call this structure the "Kremlin wall" for its resemblance to the
embattlements on top of the walls of the Moscow Kremlin.
Its period with a very good precision scales as $N^{1/2}$ and
numerically is equal to the number of electrons in the outer
crystalline row of the island. 
The amplitude of the oscillations
does not change with $N$ appreciably. 
The minima of $\delta E$ are associated with 
the highly symmetric electron configurations that can be formed at
some distinct numbers of electron $N_m$ that we call 
{\em magic numbers}. One of the examples of such configurations 
$N=85$ is displayed in Fig.~\ref{fig2}.
%---------------------------------------------------------------------------------
%
%	Fig.1_5
%
%---------------------------------------------------------------------------------

\begin{figure}
\centerline{
\psfig{file=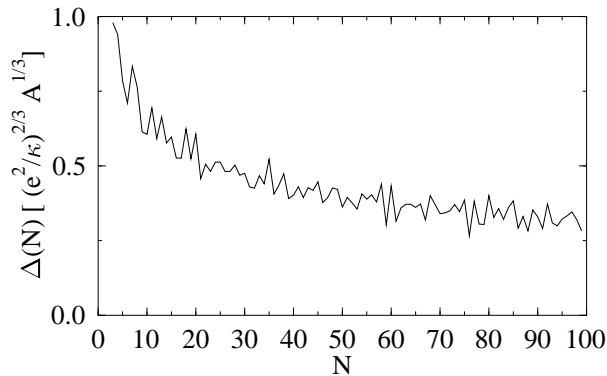,height=1.9in,bbllx=60pt,bblly=103pt,bburx=489pt,bbury=369pt}
}
\vspace{0.1in} 
\setlength{\columnwidth}{3.2in}
\centerline{\caption{
The charging energy of the Wigner crystal island as a function of the 
number of electrons in the absence of screening.
\label{fig3}
}}
\vspace{-0.1in}
\end{figure}

One can use the numerically
obtained ground state energies to calculate 
the differential charging energy $\Delta (N)$ by Eq.~(\ref{capacitance}).
The results are shown in Fig.~\ref{fig3}.
Note that although the average charging energy $\sim e^2/\kappa R$ 
decreases $\propto N^{-1/3}$ the relative fluctuations remain 
$\approx 0.15$ for number of electrons in the well
$N < 100$.
%---------------------------------------------------------------------------------
%
%	Fig.2
%
%---------------------------------------------------------------------------------

\begin{figure}
\centerline{
\psfig{file=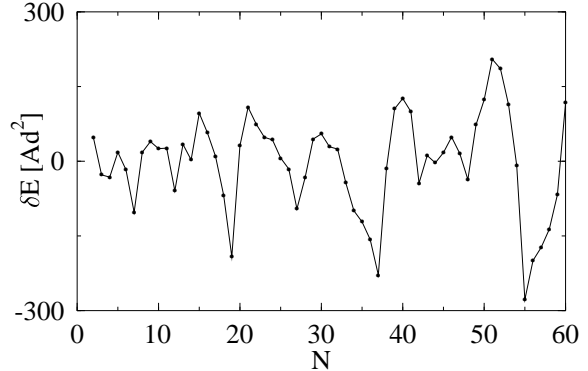,height=1.9in,bbllx=60pt,bblly=103pt,bburx=489pt,bbury=369pt}
}
\vspace{0.1in} 
\setlength{\columnwidth}{3.2in}
\centerline{\caption{
The fluctuating part of the ground state energy of the model with
the exponential interaction (\protect{\ref{Exp_int}}) for $A=10^{-8}, U_0=1, d=1$.
\label{fig4}
}}
\vspace{-0.1in}
\end{figure}

It was surprising for us to find that quasi-periodic oscillations similar to 
Fig.~\ref{fig1} exist
for all the types of interactions that we have tried. The results for
a particular example of the exponential interactions 
described by Eq.~\ref{Exp_int} are presented in Figure~\ref{fig4}.
From the total energy the smooth component of the form
$\bar{E}(N) = \eta_1 N^2 + \eta_2 N + \eta_3 N^{2/3} + \eta_4 N^{1/2} + \eta_5 N^{1/3}$
is subtracted,
where the coefficient $\eta_i$ are chosen to minimize the fluctuations.

\begin{figure}
\centerline{
\psfig{file=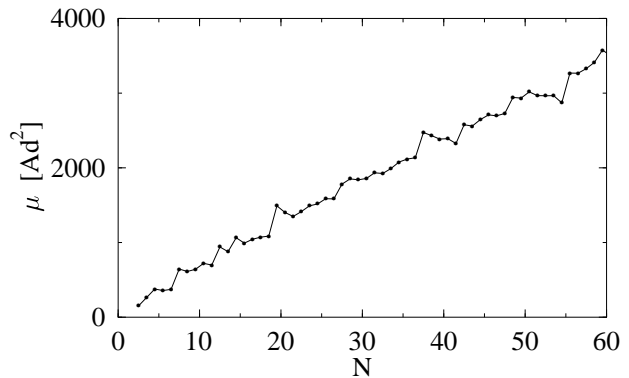,height=1.9in,bbllx=60pt,bblly=103pt,bburx=489pt,bbury=369pt}
}
\vspace{0.1in} 
\setlength{\columnwidth}{3.2in}
\centerline{\caption{
The total chemical potential $\mu(N)$ for the same parameters as in Figure 
~\protect{\ref{fig4}} (\protect{$ A=10^{-8}, U_0=1, d=1$}).
In the vicinity of 
\protect{$N = 5,7,15,20,30,40,$} etc. it is a decreasing function of number
of electrons. 
\label{fig5}
}}
\vspace{-0.1in}
\end{figure}

It is clear that the fluctuations retain their quasi-periodic form similar
to the pure Coulomb case. However, contrary to the Coulomb case they grow 
with $N$ as $N^\gamma$, where $\gamma = 0.8 \pm 0.1$.
The relative fluctuations 
of the energy also become bigger because the smooth component of the ground
state energy is significantly reduced due to screening.
As a result at some numbers of electrons in the island the fluctuations
of the charging energy exceed the average. 
The result is demonstrated in Figure ~\ref{fig5}
showing the {\em total} chemical potential of the system 
$\mu\left( N \right)=E\left( N+1\right)-E\left(N\right)$
with exponential interaction.
As one can see, at some numbers of electrons in the island the 
total chemical potential 
is a {\em decreasing} function of $N$.

%---------------------------------------------------------------------------------
%
%	Fig.2_10
%
%---------------------------------------------------------------------------------

Negative $d\mu / dN$ means that two or more electrons ``attract'' 
each other and, with increasing $\mu$ or $V_g$, 
enter the island simultaneously. In other words this means that
a few Coulomb blockade peaks merge together forming
a bunch in the charging spectrum. The detailed analysis of this instability is
given in the next Section.

%---------------------------------------------------------------------------------
%
%	Fig.2_15
%
%---------------------------------------------------------------------------------

% ---------------------------------------------------------------------------
%
%
%	Toy model
%
%
% ---------------------------------------------------------------------------

\section{A Simple Model for the Short-Range Interactions}
\label{Toy_Model}

In the previous Section we saw that bunching is associated with 
screening of the interactions in the island. Once these
interactions become short-range the fluctuations of the charging
energy exceed the average value. To understand the nature of these 
fluctuations it is instructive therefore to consider the simplest
form of the short-range interactions: the hard wall interactions.

Assume that electrons are hard disks of diameter $d$ 
put into the parabolic confinement. The energy of the system now
is given by 
\begin{equation}
E = A \sum_i \bbox{r}^2_i,
\label{model_energy}
\end{equation}
where $\bbox{r}_i$ are the coordinates of the centers of the discs.
This expression formally coincides with the moment of inertia of the system of
particles of mass $A$ relative to the center of the parabolic confinement. 
Hence, to minimize the energy one has to minimize the 
corresponding moment of inertia. As it is well known the moment of inertia 
of a system is minimum relative to the center of mass of the system.
We conclude that the center of parabola is situated in the center of mass
of the minimum energy configuration.

%---------------------------------------------------------------------------------
%
%	Fig.2_2
%
%---------------------------------------------------------------------------------
\begin{figure}
\centerline{
\psfig{file=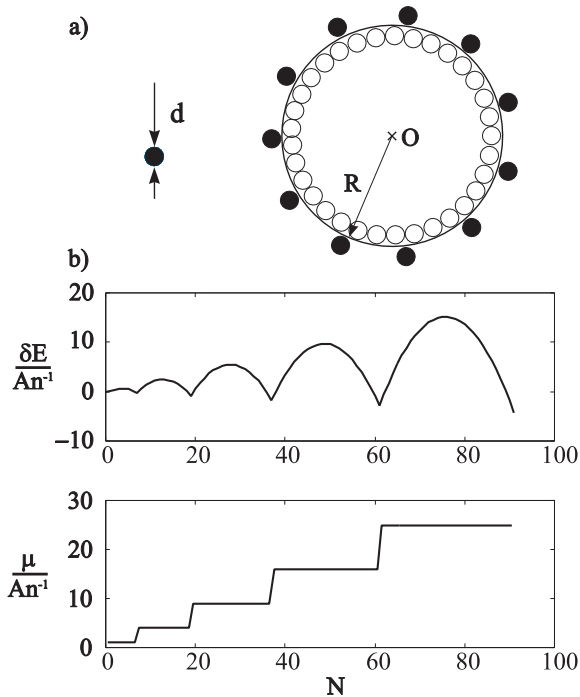,height=3.6in,bbllx=116pt,bblly=195pt,bburx=479pt,bbury=691pt}
}
\vspace{0.1in} 
\setlength{\columnwidth}{3.2in}
\centerline{\caption{
a) Disk consisting of an integer number of closed electron shells
and a few electron in the new outer shell.
One of such inner shells is shown by empty circles.  
New electrons added to the disk (dark circles) are put on
its surface. In the simplest approach the new electrons
are spread uniformly over the circumference of the disk. 
b) The fluctuating part of the total energy of the system (top)
and the total chemical potential (bottom) obtained in this approximation.
\label{fig6}
}}
\vspace{-0.1in}
\end{figure}

Assume that at some number of particles in the well they form a highly symmetric
configuration consisting of an integer number of closed radial shells.
These configurations are observed when the number of particles
corresponds to the deep minima of the fluctuating part of the 
energy of the system (see Fig.~\ref{fig1}).
The electrons in these magic number configurations form a
figure very close to the perfect hexagon. If the number of electrons
is not too big the hexagon is almost indistinguishable from 
disk. In fact the amplitude of the deviations of disk from hexagon are
less than the radius of particles if the number of particles
in the well does not exceed $175$. Below for the sake of instructiveness
we replace hexagon with disk.

When adding a new electron to the symmetric configuration it has
to be placed somewhere on the boundary of the island. The moment of inertia
of the system relative to the center of the disk is then
\begin{equation}
E_O = AN_0R^2/2 + A\delta N ( R + d\sqrt{3}/2 )^2,
\label{E_CD}
\end{equation}
where $AN_0$ and $R$ are the mass and the radius of the disk, and $\delta N$
is the number of newly added electrons (see Fig.~{\ref{fig6}}).
In the first approximation one can think that electrons
are spread uniformly over the circumference of the disk (Fig.~{\ref{fig6}}a).
Then the center of mass of the system coincides with point $O$ and its
energy is given by Eq. ~(\ref{E_CD}). It is necessary to notice now
that this expression is linear in the number of new electrons. At the
same time the average moment of inertia of the system grows $\propto N^2$.
Indeed it is equal to $N\bar{R}^2/2$, with $\bar{R}$ being the average
radius of the island determined from the expression $n \pi \bar{R}^2 = N$,
where $n=\left( d^2\sqrt{3}/2 \right)^{-1}$ is the density of
centers of particles.
Hence the expression ~(\ref{E_CD}) is just a linear approximation
to the average energy with the difference between them being the fluctuations.
The fluctuating part of the energy and the total chemical potential of the
system obtained in this approximation are shown in Fig. ~\ref{fig6}b.

The total chemical potential as a function of the number of electrons 
comprises the series of steps. This dependence resembles one
for the system of non-interacting electrons in the conditions of the
integer quantum Hall effect. 
The fluctuating part of energy is just a sequence of parabolas. 
The minima of the curve correspond to the deep minima of the Kremlin wall structure
shown in Fig. {\ref{fig1}}.

\begin{figure}
\centerline{
\psfig{file=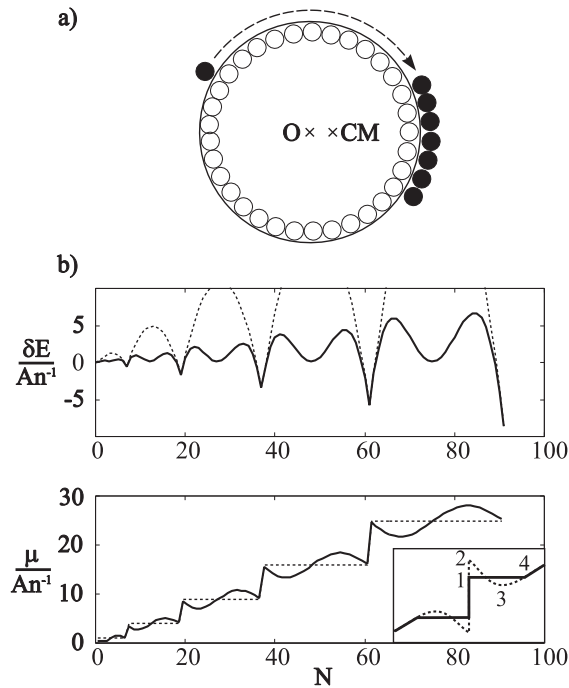,height=3.6in,bbllx=116pt,bblly=195pt,bburx=479pt,bbury=691pt}
}
\vspace{0.2in} 
\setlength{\columnwidth}{3.2in}
\centerline{\caption{
a) Clustering of electrons on one side of the disk that
shifts the position of the center of mass (point CM) from the center
of the disk, decreasing the total energy \protect{(Eq. ~(\ref{E_CM}))}.
b) The fluctuating part of the total energy (top, bold line) reduced relative
to the simplistic approach (dashed line) and the total chemical potential 
(bottom, bold line) contrasted to the simplistic approach (dashed line).
In the regions of negative $d\mu/dN$ electrons enter
the island in bunches. The number of electrons in a bunch is
determined from the Maxwell's rule, illustrated in the inset. 
\label{fig7}
}}
\vspace{-0.1in}
\end{figure}

The appearance of the sequence of shallow minima in Fig. {\ref{fig1}}  
is associated with displacements of the electrons on the surface of the disk.
One can notice that these electrons are actually attracted to each other.
Hence instead of being uniformly spread over the surface they form a 
compact cluster (see Fig. {\ref{fig7}a}). This interaction is similar 
to the attraction between two people in a boat. One person tilts the
deck of the boat and another has a tendency to slide down the slope.
For our model this can be most easily understood using the analogy with
the moment of inertia. Indeed due to the known theorem the moment
of inertia relative to point $O$ is related to one with respect to
the center of mass by
\begin{equation}
E_{\rm CM} = E_{O} - ANr_{\rm CM}^2.
\label{E_CM}
\end{equation}
Here $r_{\rm CM}$ is the distance from the center of mass to the point $O$.
It is clear that by forming a compact group the particles on the surface
do not change $E_{O}$, while at the same time they displace the center
of mass. But by doing this they decrease the total energy, 
as it follows from the previous equation. 
The results are displayed in Fig. {\ref{fig7}}.

The fluctuating part of energy now looks similar to that in Fig. {\ref{fig1}}.
A simple calculation shows that this Kremlin wall structure is described
by the following dependence:
\begin{equation}
\frac {\delta E\left( \delta N\right)}{An^{-1}} =
\frac 2{\pi} \delta N \left( \delta N_0 - \delta N \right)
- \frac {4N_0}{\pi^2} \left[ \sin \left( 
\frac {\pi \delta N}{\delta N_0} \right) \right]^2,
\label{dE_model}
\end{equation}
where $\delta N_0 = 2\pi R_0/d = \left[ \sqrt{2 \pi N_0} \right]$, is the number 
of electrons in the outer shell. By $\left[ \cdots \right]$
we denote the integer part of a number. 
The first term in this equation describes the pure shell effect
(without attraction), while the second is the result of displacement of
the center of mass (Eq. ~(\ref{E_CM})).

%---------------------------------------------------------------------------------
%
%	Fig.2_4
%
%---------------------------------------------------------------------------------

Another interesting feature of the hard-disk model is that the total
chemical potential for some fillings of the island 
is decreasing as a function of number
of electrons. In this case, if the chemical potential in the island is fixed by
an external gate, the system becomes unstable
and a few electrons enter the island simultaneously. 
This is a result
of the aforementioned attraction between electrons on the surface of the disk.
This attraction can be thought of as a multi-polaronic effect based on the spring 
connecting the magic number configuration 
with the center of the parabolic confinement. 
Therefore we call this effect the {\em confinement multi-polaronic effect}.

Let us examine this instability in more detail. 
Assume that the island is in the proximity of a metallic gate, in which
the chemical potential is equal to $eV_g$. When $\Delta N$ electrons tunnel 
from the gate to the island the energy of the system is changed by the following amount:
\begin{equation}
\Delta E = \int_{N_0}^{N_0 + \Delta N} \mu \left( N\right) dN - 
\alpha eV_g\Delta N.
\label{}
\end{equation} 
The first and the second terms in this expression describe 
the change of the total energy of the island and 
the work done by the external source respectively.
Introducing $\delta \mu \left( N \right) = \mu \left( N \right) - \alpha eV_g$,
we obtain the condition at which this tunneling is energetically possible:
\begin{equation}
\Delta E = \int_{N_0}^{N_0 + \Delta N} \delta \mu \left( N\right) dN = 0.
\label{delta_E}
\end{equation}  
This condition is satisfied e.g. when the external chemical potential reaches
the value corresponding to point $1$ in the inset of Figure \ref{fig7}b.
At this point the integral of the difference between the dashed line,
representing $\mu \left( N\right)$, and the bold line, corresponding to 
$\alpha eV_g$ is equal to zero.
This condition is similar to the Maxwell's rule well-known in
the the theory of real gases\cite{Landau96}.
Hence, instead of entering one by one (following the curve 1-2-3-4) 
a few electrons tunnel to the island simultaneously, in bunch.
This bunching is brought about by attraction between the electrons
that start forming a new shell. 
As the hard disk model is particle-hole symmetric
the bunching is also reproduced when electrons are about to finish the shell, as 
it is shown in the lower part of the inset. 

Comparison of Fig.~\ref{fig7}b with Fig.~\ref{fig4} and \ref{fig5}
shows that our model somewhat overestimates the oscillations of $E$
and $\mu$. For example our model predicts that $\delta E \propto N$,
while in the numerical experiments $\delta E \propto N^{\gamma}$, 
with $\gamma = 0.8 \pm 0.1$. This overestimation is associated with
the main assumption of this model about the existence of the perfect
disk-like shells, that brings about a strong degeneracy of the 
addition one-electron energies. 
In other words the one-electron density of states
in our model comprises the series of sharp $\delta$-function
like peaks. This is a good approximation
for small $N$, when these shells indeed exist, however in the 
form of perfect hexagons. At large $N$ ($N > 55$) the configurations with
perfect shells are never formed and the surface of the island is
always rough. The peaks in the density of states existing at small $N$
become smeared at large $N$.
Nevertheless the density of states retains periodic variations.
Hence while the variations of the ground state energy persist
their amplitude is reduced, with the exponent $\gamma$ going down 
from $1$ to $\approx 0.8$. The detailed study of these variations is a
subject of future work.

% ---------------------------------------------------------------------------
%
%
%	The Long-Range Interactions
%
%
% ---------------------------------------------------------------------------

\section{The Coulomb Interaction}
\label{Coulomb}

The universality of the observed fluctuations of energy is not
accidental. The case of the long-range interactions
can be reduced to the simple model described in the previous Section.
To do so one has to notice that
if the number of electrons in the dot is not too big 
the electrons are situated in an effective potential that is 
very close to parabolic.
This effective potential is significantly reduced with 
respect to the bare one due the to screening by the Wigner crystal:
\begin{equation}
A \rightarrow \tilde{A} \simeq  A / \left| \epsilon \left(1/R \right) \right|,
\label{renorm_curv}
\end{equation}
where $\tilde{A}$ is the renormalized external potential, 
$\epsilon \left( q \right)$ is the static dielectric function rendered by 
the Wigner crystal. In the Thomas-Fermi approximation this
dielectric function is given by
\begin{equation}
\epsilon \left( q \right) = 1 + \tilde{U}_q 
\left( d{\cal E} _{\rm corr}/dn \right) ^ {-1}.
\label{eps_wigner}
\end{equation}
Here $\tilde{U}_q=2\pi e^2/\kappa q$ is the Fourier transform of 
the bare interaction potential,
${\cal E} _{\rm corr}(n) = - \beta e^2 n^{1/2} /\kappa$ 
is the correlation energy of the Wigner lattice per particle~\cite{Bonsall77},
with $\beta$ being a numerical constant.
Putting formulas (\ref{renorm_curv}) and (\ref{eps_wigner}) together
we obtain the following estimate for the renormalized confinement:
\begin{equation}
\tilde{A} \sim A \frac {a_0}{R} \sim \frac A {\sqrt{N}}, 
\end{equation}
where $a_0 = \sqrt{2/n_0\sqrt{3}} \propto N^{-1/6}$ is the 
lattice spacing in the middle of the island, and $R \propto N^{1/3}$
is its radius (see also Appendix ~\ref{App_A}). 
Substituting $\tilde{A}$ into Eq. (\ref{dE_model}) and replacing $d$
by the characteristic lattice spacing in the island $a_0$
it is possible to obtain the uniform upper bound for the fluctuations
of the total energy in the Coulomb case:
\begin{equation}
\left| \delta E(N) \right| \lesssim 0.3 (e^2/\kappa)^{2/3} A^{1/3} N^{1/6}.
\label{prediction}
\end{equation}
The numerical results for $\delta E (N)$ presented in Fig.~\ref{fig1}
agree with the prediction of Eq.~(\ref{prediction}) in the sense that
$\delta E(N)$ has a very weak dependence on $N$.
However the amplitude of the oscillations   
observed in the numerical experiments is
by a factor of $5$ smaller than predicted by this upper bound
(see Fig.~\ref{fig1}).
This discrepancy is due to the presence of the elastic
deformations in the Coulomb Wigner crystal. In contrast
to the case of the short-range interactions due to the
smallness of the Young's modulus of the Coulomb crystal
the newly added electrons on the surface of the island
(shown by the dark circles in Fig.~\ref{fig7})
sink into the island, further decreasing the shell effect.
This effect warrants a more detailed study.

% ---------------------------------------------------------------------------
%
%	Conclusions
%
% ---------------------------------------------------------------------------

\section{Conclusions}
\label{Conclusion}
We have studied the system of classical interacting particles in the 
parabolic confinement. We have observed 
the periodic variations of the ground state energy 
that are universal, i.e. independent of the exact form of interactions.
These variations are shown to be of purely geometric origin and 
are associated with the combination of the shell effect and the polaronic effect
based on the confining parabola. 
If the electron-electron interactions in the island are short-range
these variations bring about periodic bunching in the charging spectrum.

% ---------------------------------------------------------------------------
%
%	Acknowledgements
%
% ---------------------------------------------------------------------------

\section{Acknowledgements}
\label{Acknowledgements}

We are grateful to R.~Berkovits, M.~Fogler, and N.~Zhitenev for
useful discussions. This work is supported by NSF grant DMR-9616880.

% ---------------------------------------------------------------------------
%
%	Smooth Part of the Energy of the Wigner Crystal Island
%	with Coulomb Interactions.
%
%
% ---------------------------------------------------------------------------

\appendix
\section{Smooth Part of the Energy of the Wigner Crystal Island
with Coulomb Interactions.}
\label{App_A}
In this Appendix we obtain the scaling of different contributions
to the total energy of a circular dot filled with the Wigner crystal.
In essence we derive Eq. (\ref{E_ave}). An analogous problem
was considered in literature for the system of point charges
of the surface of sphere\cite{Morris96}.

We start by naming the length scales involved. 
They are $R \propto N^{1/3}$, the radius of the island
given by Eq.~(\ref{n_r}), and the lattice spacing in the center
of the island $a_0 = \sqrt{2/n_0\sqrt{3}} \propto N^{-1/6}$ 
(see also Eq. (\ref{n_r})).

The expansion of the total energy into the series of powers of $N$ can
be written in the following form:
\begin{equation}
E_{\rm tot} = E_{\rm ES} + E_{\rm corr} + E_{\rm OS} + E_{\rm sur} + \ldots.
\label{General_expansion}
\end{equation}
The various terms in this expression are respectively:
the electrostatic, the correlation, the overscreening, and the surface energies.
The electrostatic energy can be calculated from the solution of the electrostatic 
problem  (\ref{n_r}):
\begin{equation}
E_{\rm ES} = \frac 12 \int_0^R 2\pi r dr n\left( r \right) \left( \phi_0 + Ar^2 \right),
\end{equation}
where $\phi_0=2AR^2$ is the total potential inside the island.
As a result we obtain
\begin{equation}
E_{\rm ES} = \frac 65 NAR^2 \propto N^{5/3}.
\label{E_es}
\end{equation}

The correlation energy describes lowering of the energy due to the
transition from liquid, assumed in the electrostatic model, to crystal.
In case of a big island the energy reduction can be calculated 
in the local density approximation from the known correlation energy
of the uniform Wigner crystal~\cite{Bonsall77}:
\begin{equation}
\begin{array}{ll}
{\displaystyle
E_{\rm corr} }&
{\displaystyle
= -\int_0^R 2 \pi r dr n\left( r \right) \beta e^2 \sqrt{n\left( r\right)} 
/\kappa } \\ \\
&
{\displaystyle
= -\frac {12\beta N}{7\pi} \sqrt{\frac {A R e^2}{\kappa}} \propto N^{7/6},   }
\label{E_corr}
\end{array}
\end{equation}
where the coefficient $\beta$ was calculated in Ref.\onlinecite{Bonsall77}.
For the triangular lattice it is equal to $\beta_{\triangle}=1.960517$,
for the square lattice $\beta_{\Box}=1.949299$. The actually
observed value of this coefficient $\beta = 1.956$ lies between these
two values.

%---------------------------------------------------------------------------------
%
%	Fig.4
%
%---------------------------------------------------------------------------------
\begin{figure}
\centerline{
\psfig{file=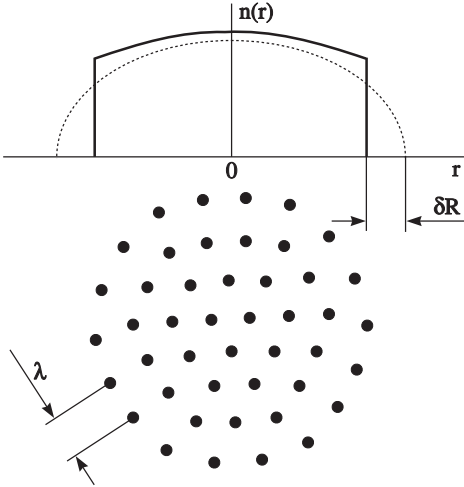,height=2.3in,bbllx=180pt,bblly=222pt,bburx=578pt,bbury=567pt}
}
\vspace{0.1in} 
\setlength{\columnwidth}{3.2in}
\centerline{\caption{
The density profile in a small crystalline sample (solid line).
It is quite different from the solution of the electrostatic problem
(dashed line) valid in the limit of very large island. 
The real density looks like a box if the
sample contains only few crystalline rows in it.
As an example on the bottom we show the ground state configuration
of $N=45$ electrons.
\label{fig8}
}}
\vspace{-0.1in}
\end{figure}

The next term in the energy expansion takes into account the deviation
of the density profile of the crystalline island from the semicircle
(\ref{n_r}) obtained in the electrostatic model.
Due to the screening the variations of the average density of crystal are
slightly bigger than obtained in electrostatics.
Note that in case of electron liquid these variations are weaker than
the electrostatic ones.
To emphasize this difference between liquid and crystal we call the
corresponding correction to energy the overscreening energy.
To estimate the correction we notice that the average density of the 
crystal deviates from the electrostatic density by $\delta n / n \sim - r_s / R$,
where $r_s$ is the screening radius of the Wigner crystal equal to the 
negative lattice spacing: $r_s \sim - a_0$ (Ref. \onlinecite{Efros}). 
As a result the distances
between electrons change by the same factor and the correlation 
energy acquires the following correction:
\begin{equation}
E_{\rm OS} \sim E_{\rm corr} a_0/R 
\sim \left( e^2/\kappa \right)^{2/3} A^{1/3} N^{2/3}.
\label{E_os}
\end{equation}

The last contribution to the total energy is the surface energy. 
It describes the 
deviations of the correlation energy of electrons on the
surface from that in the bulk. 
To find this deviation it is essential to know 
what is the lattice spacing at the edge of the island $\lambda$
(see Fig. ~\ref{fig8}).
Below we address this problem in some detail.

The electrostatic formula (\ref{n_r}) is 
correct on the scales significantly exceeding the screening radius.
As it was pointed out before  the screening radius 
of the Wigner crystal is of the
order of its lattice constant~\cite{Efros}. 
Hence the electrostatic formula breaks down at the
distances from the edge $\delta R$
of the order of the lattice spacing on the surface (see Fig. ~\ref{fig8}):
\begin{equation}
\delta R \sim \lambda.
\end{equation}
Matching the electrostatic density of 
electrons at this distance
with the density of the Wigner crystal at the edge
we obtain:
\begin{equation}
n\left( R-\lambda \right) = n_0 \sqrt{\frac {\lambda}R } 
\sim \frac 1{\lambda ^2},
\label{equation_for_lambda}
\end{equation}
and
\begin{equation}
\lambda \sim a_0^{4/5}R^{1/5} \propto N^{-1/15}.
\label{lambda_par}
\end{equation}
Our numerical data agree very well with this theory and give the following 
coefficients:
\begin{equation}
\begin{array}{c}
{\lambda = 0.88 a_0^{4/5}R^{1/5},} \\ \\
{\delta R = 0.76 a_0^{4/5}R^{1/5}.}
\end{array}
\label{lambda}
\end{equation}
Notice that the inter-electron density at the edge
differs very slightly from the lattice constant in the
center of the sample if radius of the island is not too big.
Hence the concentration of electrons in this case also
does not change too much. 
This is displayed by a box like density profile in Fig.~\ref{fig8}.

The surface energy can now easily be estimated.
It is of the order of the correlation energy per particle on
the surface times the total number of particles there $R/\lambda$:
\begin{equation}
E_{\rm surf} \sim \frac {e^2}{\kappa \lambda} \frac {R}{\lambda}
\sim \left( e^2/\kappa \right)^{2/3} A^{1/3} N^{7/15}.
\label{E_surf}
\end{equation}

Putting Equations (\ref{E_es}) \ldots (\ref{E_surf}) together we obtain
the expansion (\ref{E_ave}).

\vspace{-0.2in}
%-------------------------------------------------------------------------------

\end{multicols}

\begin{references}
\vspace{-0.4in}

\bibitem{AshooriNew} N.~B.~Zhitenev, R.~C.~Ashoori, L.~N.~Pfeiffer, 
and K.~W.~West, preprint cond-mat/9703241.

\bibitem{AshooriOld} R.~C.~Ashoori, H.~L.~Stormer, J.~C.~Weiner, L.~N.~Pfeiffer,
K.~W.~Baldwin, S.~J.~Pearton, and K.~W.~West, \prl {\bf 68}, 3088 (1992); 
R.~C.~Ashoori, H.~L.~Stormer, J.~C.~Weiner, L.~N.~Pfeiffer,
K.~W.~Baldwin, and K.~W.~West, \prl {\bf 71}, 613 (1993).

\bibitem{Phylipps} Y.~Wan, G.~Ortiz, and P.~Phyllips, \prl 75, 2879 (1995).

\bibitem{Raikh} M.~E.~Raikh, L.~I.~Glazman, L.~E.~Zhukov, \prl, {\bf 77}, 1354 (1996).

\bibitem{Peeters94} V.~M.~Bedanov and F.~M.~Petters, \prb {\bf 49}, 2667 (1994).

\bibitem{Sneddon66} Ian N. Sneddon, {\em Mixed Boundary Value Problems 
in Potential Theory} (John Wiley \& Sons, Inc., New York, 1966).

\bibitem{Landau96} L.~D.~Landau and E.~M.~Lifshitz, {\em Statistical Physics, 
part 1} (Reed Educational and Publishing Ltd, Oxford, 1980), p. 261.

\bibitem{Morris96} J.~R.~Morris, D.~M.~Deaven, K.~M.~Ho, \prb {\bf 53}, R1740 (1996).

\bibitem{Bonsall77} L.~Bonsall, A.~A.~Maradudin, \prb {\bf 15}, 1959 (1977).

\bibitem{Efros} A.~L.~Efros, Solid State Communications {\bf 65}, 1281 (1988).

%\bibitem{} .
%\bibitem{} .
%\bibitem{} .
%\bibitem{} .
%\bibitem{} .

\end{references}
\end{document}